\documentclass [useAMS]{mn2e}
\usepackage {graphicx}

\begin{document}

\title{An analytic model of rotating hot spot and kHz QPOs in X-ray binaries}
\date{Accepted }
\author[]{Ding-Xiong Wang$^{1,3}$, Ren-Yi Ma$^1$, Wei-Hua Lei$^1$ and Guo-Zheng Yao$^2$ \\
$1$ Department of Physics, Huazhong University of Science and Technology, Wuhan,430074,China \\
$2$ Department of Physics, Beijing Normal University, Beijing 100875, China \\
$3$ Send offprint requests to: D.-X. Wang (dxwang@hust.edu.cn) }
\maketitle

\begin{abstract}

An analytic model of rotating hot spot is proposed to explain
kilohertz quasi-periodic oscillations (kHz QPO) in X-ray binaries,
which is based on the magnetic coupling (MC) of a rotating black
hole (BH) with its surrounding accretion disc. The hot spot in the
inner region of the disc is produced by energy transferred from a
spinning BH with non-axisymmetric magnetic field. The frequency
and energy band of the hot spot turns out to be related to six
parameters, by which the strength and the azimuthal angular region
of the bulging magnetic field on the BH horizon, the mass and spin
of the BH, and the power-law index of the magnetic field varying
with the radial coordinate on the disc are described. In addition,
the correlation of the fluctuation of the bulging magnetic field
with the widths of QPO frequency is discussed.

\end{abstract}


\section{INTRODUCTION}

With the existence of a magnetic field connecting a Kerr black
hole (BH) to its surrounding disc, energy and angular momentum can
be transferred from the BH to the disc (Blandford 1999; Li 2000;
Li 2002a, hereafter Li02a; Wang, Xiao {\&} Lei 2002, hereafter
WXL), which can be regarded as one of the variants of the
Blandford-Znajek (BZ) process proposed two decades ago (Blandford
{\&} Znajek 1977). Henceforth this energy mechanism is referred to
as the magnetic coupling (MC) process. Recently Wilms et al.
(2001) found that \textit{XMM-Newton} observation of the nearby
bright Seyfert 1 galaxy MCG-6-30-15 reveals an extremely broad and
redshifted $Fe\mbox{ }K\alpha $ line indicating its origin from
the very most central region of the accretion disc, and they
suggested that the rotating energy of a BH is extracted by the
magnetic field connecting the BH or plunging region to the disc.
Later Li pointed out that the magnetic coupling of a spinning BH
with a disc can produce a very steep emissivity with index $\alpha
= 4.3 \sim 5.0$, which is consistent with the above
\textit{XMM-Newton} observation (Li 2002b, hereafter Li02b). This
result can be regarded as the observation signatures of the
existence of the MC process. Very recently we obtained the same
result in a more detailed MC model, which is consistent with the
\textit{XMM-Newton} observation with reference to a variety of
parameters of the BH-disc system (Wang, Lei {\&} Ma 2003,
hereafter WLM).

Another relevance of MC model to the observation is kilohertz
quasi-periodic oscillations (kHz QPO) in X-ray binaries. As argued
by van der Klis (2000), kHz QPOs in X-ray binaries probably
originate from the inner edge of an accretion disc with a BH of
stellar-mass, since millisecond is the natural timescale for
accretion process in these regions. Recently Li suggested that the
non-axisymmetric MC of a spinning BH of stellar-mass with a disc
might be used to explain the kHz QPOs (Li 2001, hereafter Li01).
Motivated by Li's suggestion we propose an analytic model of a
rotating hot spot to explain the kHz QPOs, where the hot spot
arises from the energy transferred to the inner region of the
relativistic disc by the closed field lines of a non-axisymmetric
magnetic field on the BH horizon. Six parameters are involved in
our model. Among them three are used to describe the strength and
azimuthal angular region of the bulging magnetic field on the BH
horizon, another two are used to describe mass and spin of the
Kerr BH, and one is the power-law index of the magnetic field
varying with the radial coordinate on the disc. We give a detailed
discussion on a rotating hot spot for fitting kHz QPOs in terms of
these parameters.

\begin{figure}
\vspace{0.5cm}
\begin{center}
\includegraphics[width=6cm]{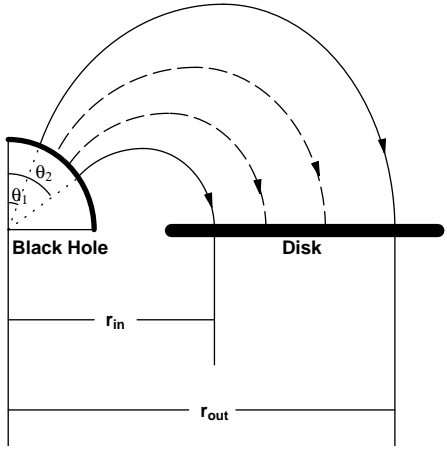}
\caption{The poloidal magnetic field connecting a spinning BH with
its surrounding disc} \label{fig1}
\end{center}
\end{figure}


\begin{figure}
\vspace{0.5cm}
\begin{center}
{\includegraphics[width=6cm]{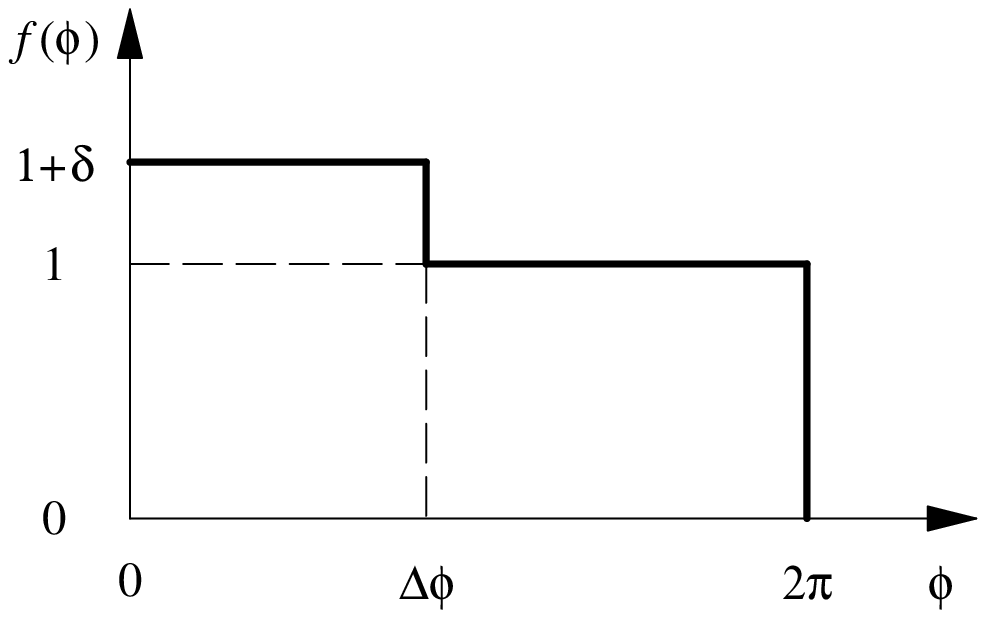}
 \centerline{(a)}
 \includegraphics[width=6cm]{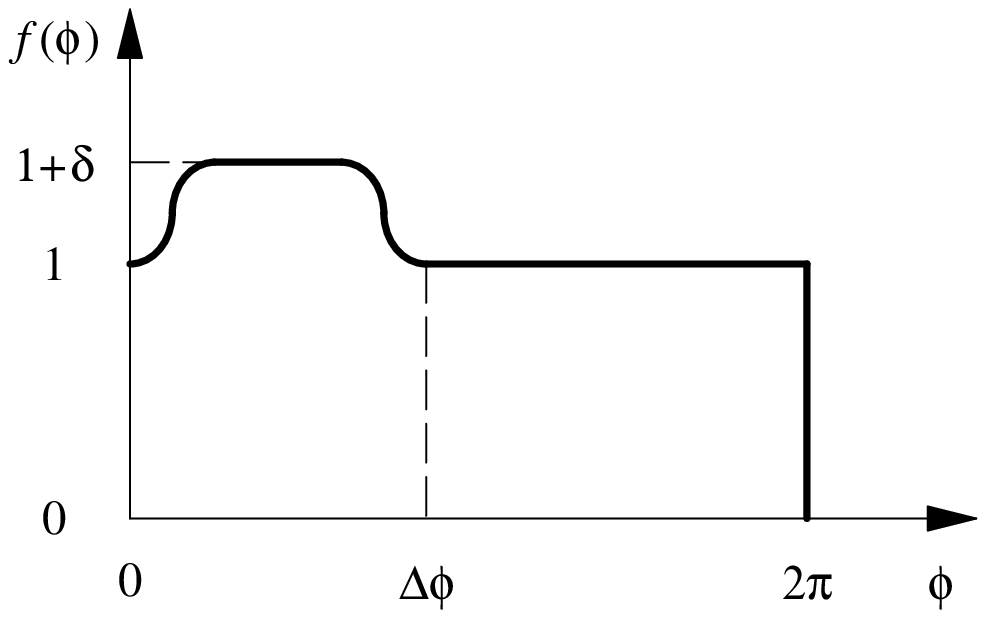}
 \centerline{(b)}}
 \caption{Azimuthal profile of the non-axisymmetric magnetic
field on the BH horizon}
\end{center}
\end{figure}

This paper is organized as follows. In Section 2 we show that the
position of a hot spot is determined by non-axisymmetric magnetic
field with a mapping relation between the poloidal angular
coordinate on the horizon and the radial coordinate on the disc.
It turns out that the hot spot is limited very close to the inner
edge of the disc for the appropriate values of the BH spin and the
power-law index of the magnetic field on the disc. In Section 3 we
prove that the hot spot can reach the energy level of emitting
X-ray, provided that the bulging magnetic field on the horizon is
strong enough to reach $ \sim 10^8Gauss$. Furthermore the rotating
frequency of a hot spot is expressed in terms of Keplerian angular
velocity at the place where $r^2F$ is maximized, where $F$ is the
sum of radiation flux due to disc accretion and that arising from
the MC between the spinning BH and the disc. In Section 4 we fit
kHz QPO widths by assuming the fluctuation of the bulging magnetic
field on the horizon. Finally, in Section 5, we give a summary for
our model.

Throughout this paper the geometric units $G = c = 1$ are used.


\section{ RADIAL POSITION OF ROTATING HOT SPOT }

In order to facilitate the discussion of the rotating hot spot we
make the following assumptions.

(i) The magnetic field is assumed to connect the BH with the
surrounding disc as shown in Fig.1. The radii $r_{in} $ and
$r_{out} $ are the radii of inner and outer boundaries of the MC
region, respectively, and $\theta _1 $ and $\theta _2 $ are the
corresponding poloidal angular coordinates on the horizon.

(ii) The toroidal profile of the magnetic field is
non-axisymmetric on the BH horizon, being expressed by a function
$f\left( \phi \right)$ of the azimuthal coordinate $\phi $ as
shown in Fig.2. The magnetic field is assumed to vary in a
power-law as the radial coordinate of the disc as suggested by
Blandford (1976).

(iii) The disc is both stable and perfectly conducting, and the
closed magnetic field lines are frozen in the disc.

(iv) The disc is thin and Keplerian, locating in the equatorial
plane of the BH with the inner boundary at the last stable
circular orbit.

Generally speaking, the magnetic field on the horizon is
non-axisymmetric. Unfortunately, a reliable profile of the
magnetic field on the horizon has not been given due to lack of
knowledge about it. As a simple analysis we assume the profile
function $f\left( \phi \right)$ as shown in Fig.2(a), which is
expressed by

\begin{equation}
\begin{array}{l}
\label{eq1} B_H \left( \phi \right) = \sqrt {\left\langle {B_H^2 }
\right\rangle } f\left( \phi \right), \\ \\
f\left( \phi \right)
\equiv \left\{ {\begin{array}{l}
 1 + \delta ,\quad\mbox{ }0 < \phi < \Delta \phi ; \\
\quad 1,\quad\quad\mbox{ }\Delta \phi \le \phi \le 2\pi . \\
 \end{array}} \right.
 \end{array}
\end{equation}

\noindent where $\sqrt {\left\langle {B_H^2 } \right\rangle } $ is
root-mean-square of the magnetic field over the poloidal angular
coordinate from $\theta _1 $ to $\theta _2 $. As shown in
Fig.2(a), the parameter $\delta $ is used to describe the strength
of the bulging magnetic field in the region $0 < \phi < \Delta
\phi $, which is the azimuthal angular range of the hot spot on
the disc. The drawback of the pattern given in Fig.2(a) lies in
the uncontinuity of the magnetic field at $\phi = 0$ and$\mbox{
}\Delta \phi $. In fact, function $f\left( \phi \right)$ can be
modified by a smooth joint-function as shown in Fig.2(b), and we
shall modify the model of rotating hot spot by using a more
realistic profile function in future.

The radial range of the hot spot is determined by a mapping
relation between the angular coordinate on the BH horizon and the
radial coordinate on the disc as follows:

\begin{equation}
\label{eq2} \cos \theta = \int_1^\xi {G\left( {a_ * ;\xi ,n}
\right)} d\xi ,
\end{equation}

\noindent where

\begin{equation}
\label{eq3} \begin{array}{l} G\left( {a_ * ;\xi ,n} \right) = \\
\\
\quad\quad \frac{\xi ^{1 - n}\chi _{ms}^2 \sqrt {1 + a_ * ^2 \chi
_{ms}^{ - 4} \xi ^{ - 2} + 2a_ * ^2 \chi _{ms}^{ - 6} \xi ^{ - 3}}
}{2\sqrt {\left( {1 + a_
* ^2 \chi _{ms}^{ - 4} + 2a_ * ^2 \chi _{ms}^{ - 6} }
\right)\left( {1 - 2\chi _{ms}^{ - 2} \xi ^{ - 1} + a_ * ^2 \chi
_{ms}^{ - 4} \xi ^{ - 2}} \right)} }.
\end{array}
\end{equation}

\begin{figure*}
\begin{center}
{\includegraphics[width=5.6cm]{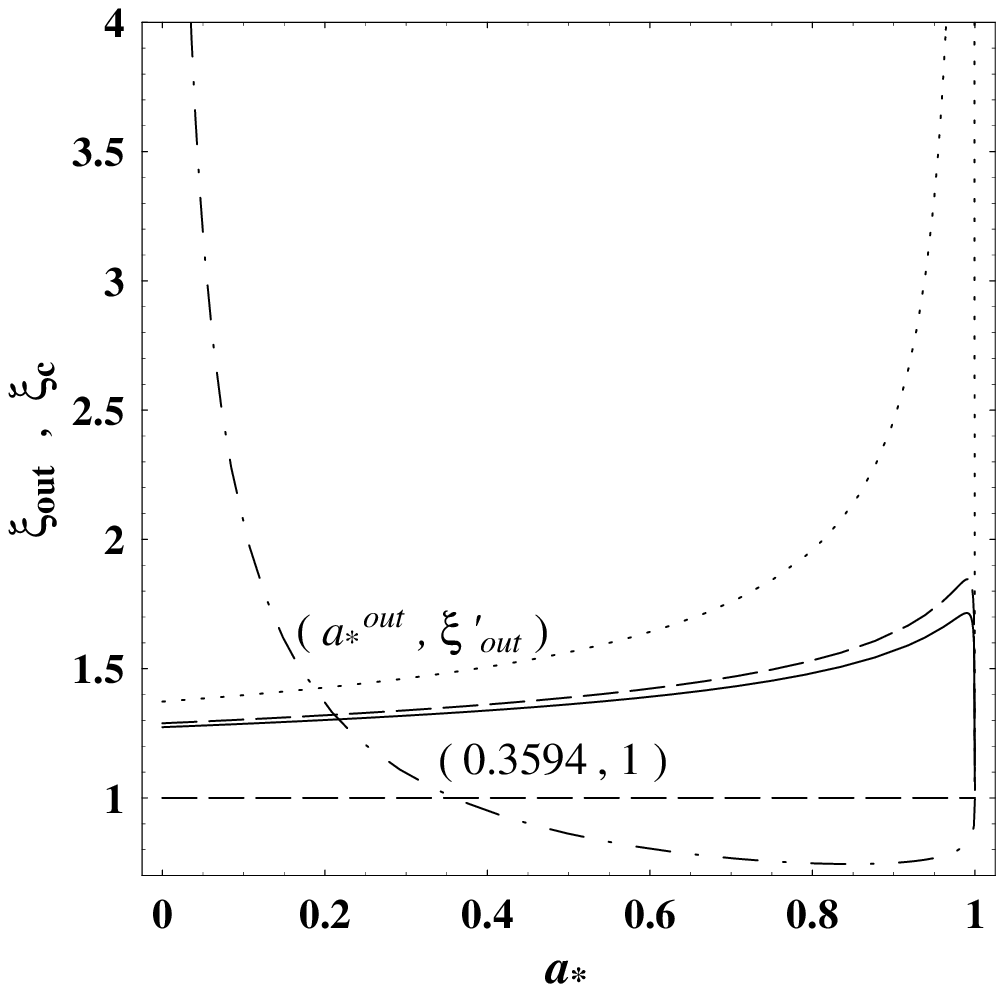} \hfill
\includegraphics[width=5.6cm]{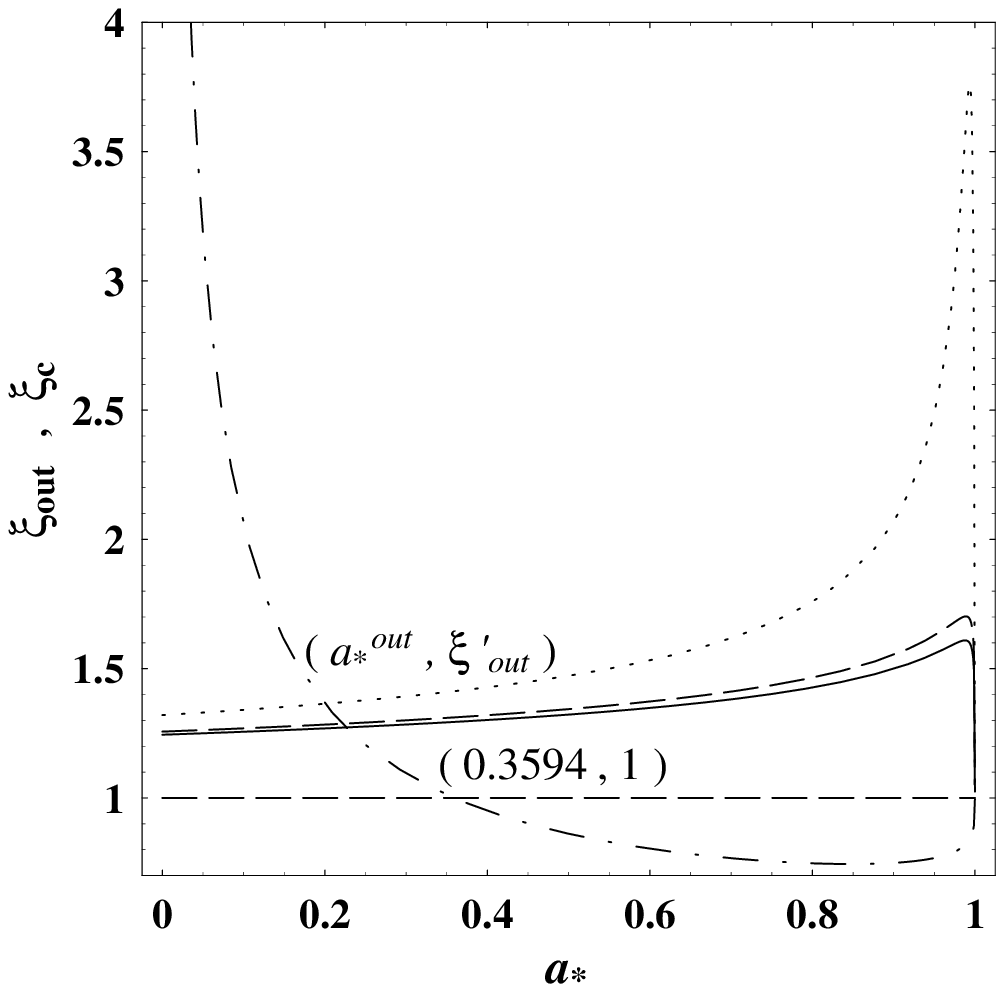} \hfill
\includegraphics[width=5.6cm]{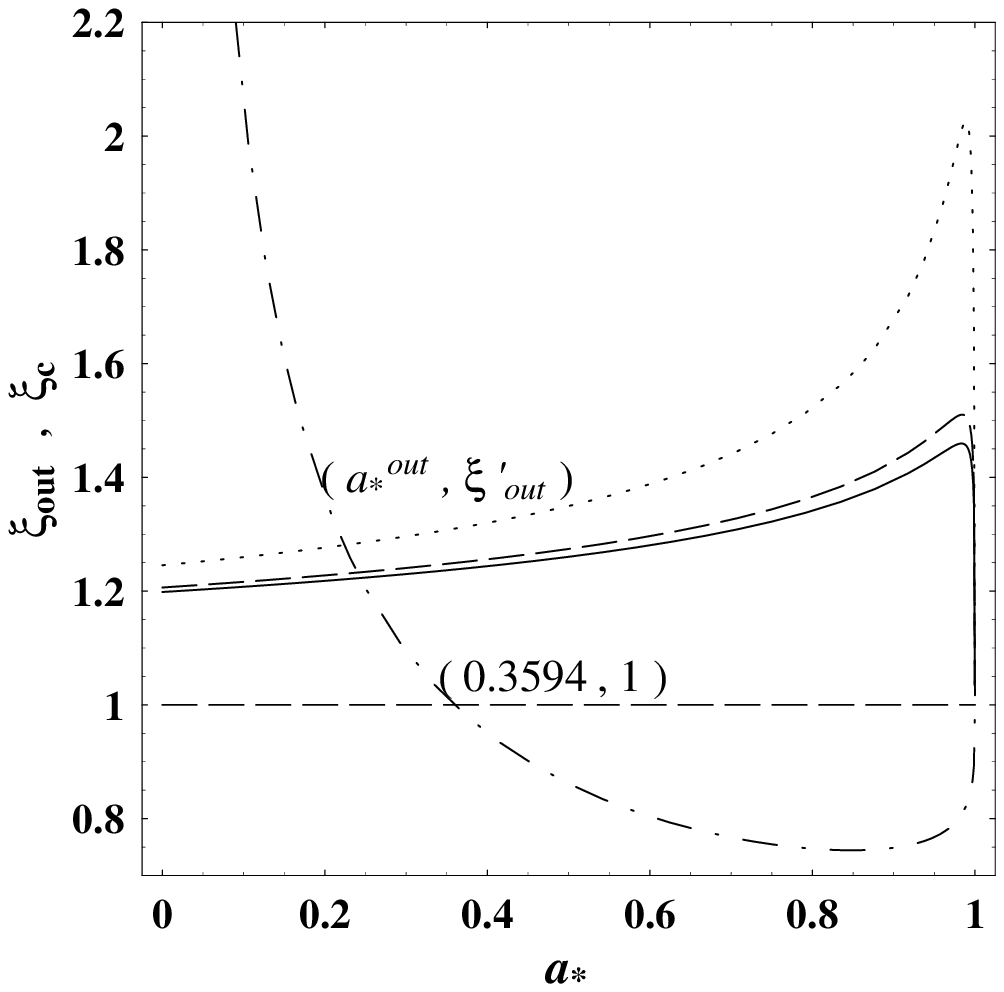}
\centerline{\hspace{0.6cm}(a)\hspace{5.65cm}(b)\hspace{5.7cm}(c)}
} \caption{The curves of $\xi_c$(dot-dashed line) and $\xi_{out}$
versus $a_*$ for $0<a_*<1$ with $n=1.1,1.5$ and 3 in solid, dashed
and dotted lines, respectively. The values of $\theta_1$ are given
as (a)$\theta_1=\pi/12$, (b)$\theta_1=\pi/6$ and
(c)$\theta_1=\pi/4$.} \label{fig2}
\end{center}
\end{figure*}

The angle $\theta _2 = \pi \mathord{\left/ {\vphantom {\pi 2}}
\right. \kern-\nulldelimiterspace} 2$ is assumed in derivation.
Equation (\ref{eq2}) associated with equation (\ref{eq3}) is the
revised version of the mapping relation derived in WXL. Parameter
$\xi \equiv r \mathord{\left/ {\vphantom {r {r_{ms} }}} \right.
\kern-\nulldelimiterspace} {r_{ms} }$ is a dimensionless radius of
the disc in terms of the radius of the last stable circular orbit,
$r_{ms} = M\chi _{ms}^2 $, and $a_ * \equiv J \mathord{\left/
{\vphantom {J {M^2}}} \right. \kern-\nulldelimiterspace} {M^2}$ is
dimensionless parameter of the BH spin defined by the BH mass $M$
and angular momentum $J$. Following Blandford (1976) the normal
components of magnetic field, $B_z $, on the disc is assumed to
vary with $\xi $ as follows.

\begin{equation}
\label{eq4} B_z \propto \xi ^{ - n}.
\end{equation}

\noindent In deriving equations (\ref{eq2}) and (\ref{eq3}) the
boundary condition is used as given in WLM, i.e.

\begin{equation}
\label{eq5} 2\pi r_H B_ \bot = 2\pi \varpi _D \left( {r_{ms} }
\right)B_z \left( {r_{ms} } \right),
\end{equation}

\noindent where $B_ \bot $ is the normal component of magnetic
field at horizon, and $\varpi _D \left( {r_{ms} } \right)$ is the
cylindrical radius at $r_{ms} $,

\begin{equation}
\label{eq6} \varpi _D \left( {r_{ms} } \right) = M\chi _{ms}^2
\sqrt {1 + \chi _{ms}^{ - 4} a_ * ^2 + 2\chi _{ms}^{ - 6} a_ * ^2
} .
\end{equation}

Equation (\ref{eq5}) implies the conservation of magnetic flux
corresponding to the two loops of the same width, and the ratio of
$B_ \bot $ to $B_z \left( {r_{ms} } \right)$ varies from 1.8 to 3
for $0 < a_ * < 1$. Considering that the strength of the magnetic
field at the horizon is likely greater than that in the disc by
numerical simulation (Ghosh {\&} Abramowicz 1997 and the
references therein), we think that the relation (\ref{eq5}) is
more reasonable than the boundary condition $B_ \bot = B_z \left(
{r_{ms} } \right)$ given in WXL.

Substituting $\theta = \theta _1 $ and $\xi = \xi _{out} $ into
equation (\ref{eq2}), we obtain an integral equation for
determining the outer boundary $\xi _{out} \equiv {r_{out} }
\mathord{\left/ {\vphantom {{r_{out} } {r_{ms} }}} \right.
\kern-\nulldelimiterspace} {r_{ms} }$ of the MC region as follows:

\begin{equation}
\label{eq7} \cos \theta _1 = \int_1^{\xi _{out} } {G\left( {a_ *
;\xi ,n} \right)} d\xi .
\end{equation}

\noindent From equation (\ref{eq7}) we find that $\xi _{out}
\left( {a_
* ;n,\theta _1 } \right)$ behaves as a non-monotonic function of
$a_
* $ for different values of $n$ and $\theta _1 $ as shown in
Fig.3.

In WLM the transfer direction of energy and angular momentum
between a rotating BH and its surrounding disc is discussed in
detail by using the co-rotation radius $r_c $, which is defined as
the radius where the angular velocity $\Omega _D $ of the disc is
equal to the BH angular velocity $\Omega _H $. The parameter $\xi
_c \left( {a_ * } \right) \equiv {r_c } \mathord{\left/ {\vphantom
{{r_c } {r_{ms} }}} \right. \kern-\nulldelimiterspace} {r_{ms} }$
is a function of $a_ * $ determined by the following equation,

\begin{equation}
\label{eq8} {\Omega _D } \mathord{\left/ {\vphantom {{\Omega _D }
{\Omega _H }}} \right. \kern-\nulldelimiterspace} {\Omega _H } =
\frac{2\left( {1 + q} \right)}{a_
* }\left[ {\left( {\sqrt \xi \chi _{ms} } \right)^3 + a_ * } \right]^{ - 1}
= 1,
\end{equation}

\noindent where

\begin{equation}
\begin{array}{l}
\label{eq9} \Omega _D = \frac{1}{M\left( {\chi ^3 + a_ * }
\right)}, \quad \Omega _H = \frac{a_ * }{2M\left( {1 + q}
\right)},\\ \\ q = \sqrt {1 - a_ * ^2 } .
\end{array}
\end{equation}

\noindent In equation (\ref{eq9}) $\chi \equiv \sqrt {r
\mathord{\left/ {\vphantom {r M}} \right.
\kern-\nulldelimiterspace} M} = \sqrt \xi \chi _{ms} $ is a
dimensionless radial parameter on the disc, and $\chi _{ms} $ is
defined as $\chi _{ms} \equiv \sqrt {{r_{ms} } \mathord{\left/
{\vphantom {{r_{ms} } M}} \right. \kern-\nulldelimiterspace} M} $.
The MC region is divided by $r_c $ into two parts: the inner MC
region (henceforth IMCR) for $1 < \xi < \xi _c $ and the outer MC
region (henceforth OMCR) for $\xi _c < \xi < \xi _{out} $. The
parameter $\xi _c $ decreasing monotonically with $a_ * $ is shown
by the dot-dashed line in Fig.3, and the transfer direction of
energy and angular momentum is described as follows.

(i) For $0.3594 < a_ * < 1$ the transfer direction is from the BH
to the disc with $\xi _c < 1$.

(ii) For $0 \le a_ * < a_ * ^{out} $ the transfer direction is
from the disc to the BH with $\xi _c > \xi _{out} $, where $a_ *
^{out} $ is the BH spin corresponding to the intersection of the
curve $\xi _c \left( {a_ * } \right)$ with the curve $\xi _{out}
\left( {a_ * ,n,\theta _1 } \right)$ as shown in Fig.3.

(iii) For $a_ * ^{out} < a_ * < 0.3594$ the transfer direction is
from the BH to OMCR with $\xi _c < \xi < \xi _{out} $, while it is
from IMCR to the BH with $1 < \xi < \xi _c $.

Since a hot spot is produced by the energy transferred from the BH
to the disc, its radial range is expressed by

\begin{equation}
\label{eq10} \left\{ {\begin{array}{l}
 \xi _c < \xi < \xi _{out} ,\mbox{ }for\mbox{ }a_ * ^{out} < a_ * <
0.3594\mbox{ }\left( {case I} \right)\mbox{ } \\
 1 < \xi < \xi _{out} ,\mbox{ }for\mbox{ }0.3594 < a_ * < 1\mbox{ }\left(
{case II} \right). \\
 \end{array}} \right.
\end{equation}

From Fig.3 we find that the outer boundary $\xi _{out} $ is very
close to $\xi _{in} = 1$ in both case $I$ and case \textit{II},
provided that the power-law index $n$ is small, such as $n =
1.1,\mbox{ 1.5}$. This result seems insensitive to the values of
$\theta _1 $. Therefore the hot spot is confined to a small radial
region very near to the inner edge of the disc.

\section{ RADIATION ENERGY AND ROTATING FREQUENCY OF ROTATING HOT
SPOT}

Based on conservation of energy and angular momentum the basic
equations for BH evolution in the coexistence of disc accretion
and the MC process are written as

\begin{equation}
\label{eq11} {dM} \mathord{\left/ {\vphantom {{dM} {dt}}} \right.
\kern-\nulldelimiterspace} {dt} = E_{ms} \dot {M}_D - P_{MC} ,
\end{equation}

\begin{equation}
\label{eq12} {dJ} \mathord{\left/ {\vphantom {{dJ} {dt}}} \right.
\kern-\nulldelimiterspace} {dt} = L_{ms} \dot {M}_D - T_{MC} .
\end{equation}

\noindent where $P_{MC} $ and $T_{MC} $ are MC power and torque in
the MC process, respectively. In WXL we derived $P_{MC} $ and
$T_{MC} $ for axisymmetric magnetic field, by using an improved
equivalent circuit as follows:

\begin{equation}
\label{eq13} \begin{array}{l}P_{MC}^{axis} = 2\left\langle {B_H^2
} \right\rangle M^2a_ * ^2 \int_1^\xi {\frac{\beta \left( {1 -
\beta } \right)G\left( {a_ * ;{\xi }',n} \right)}{2\csc ^2\theta -
\left( {1 - q} \right)}} d{\xi }',
\end{array}
\end{equation}

\begin{equation}
\label{eq14}
\begin{array}{l}
T_{MC}^{axis} = 4\left\langle {B_H^2 } \right\rangle
M^3a_ * \left( {1 + q} \right)\times \\ \\
\quad\quad\quad\quad\quad\quad\quad\quad\quad\quad\int_1^\xi
{\frac{\left( {1 - \beta } \right)G\left( {a_ * ;{\xi }',n}
\right)}{2\csc ^2\theta - \left( {1 - q} \right)}} d{\xi }'.
\end{array}
\end{equation}

\noindent Equations (\ref{eq13}) and (\ref{eq14}) can be extended
to the case for non-axisymmetric magnetic field as follows:

\begin{equation}
\label{eq15} P_{MC}^{naxis} = \lambda P_{MC}^{axis} ,
\end{equation}

\begin{equation}
\label{eq16} T_{MC}^{naxis} = \lambda T_{MC}^{axis} ,
\end{equation}

\noindent where the superscripts `\textit{axis}' and
`\textit{naxis}' indicate the quantities for axisymmetric and
non-axisymmetric magnetic field, respectively. The parameter
$\lambda $ is expressed by

\begin{equation}
\label{eq17} \lambda = \left[ {\left( {1 + \delta }
\right)\varepsilon + \left( {1 - \varepsilon } \right)} \right]^2
= \left( {1 + \delta \varepsilon } \right)^2,
\end{equation}

\noindent where ${\varepsilon \equiv \Delta \phi } \mathord{\left/
{\vphantom {{\varepsilon \equiv \Delta \phi } {2\pi }}} \right.
\kern-\nulldelimiterspace} {2\pi }$ is a parameter indicating the
azimuthal region of the bulging magnetic field on the horizon
expressed by equation (\ref{eq1}). Equations (\ref{eq15}) and
(\ref{eq16}) can be derived in the same way as in WXL, provided
that the magnetic flux between two adjacent magnetic surfaces
$\Delta \Psi = B_H 2\pi \varpi \Delta l$ is replaced by

\begin{equation}
\label{eq18} \begin{array}{l}
 \Delta \Psi = \sqrt {\left\langle
{B_H^2 } \right\rangle } \varpi \left[ {\left( {1 + \delta }
\right)\Delta \phi + \left( {2\pi - \Delta \phi } \right)}
\right]\Delta l \\ \\
\quad\quad= \sqrt {\left\langle {B_H^2 } \right\rangle } 2\pi
\varpi \left( {1 + \delta \varepsilon } \right)^2\Delta l,
\end{array}
\end{equation}

\noindent where equation (\ref{eq18}) is given for the bulging
magnetic field expressed by equation (\ref{eq1}). Obviously,
equations (\ref{eq15}) and (\ref{eq16}) reduce to axisymmetric
ones for either $\delta \to 0$ or $\varepsilon \to 0$ with
$\lambda = 1$, and they also become axisymmetric ones for
$\varepsilon \to 1$ with $\lambda = \left( {1 + \delta }
\right)^2$. Based on equations (\ref{eq13})---(\ref{eq16}) we can
express the rate of extracting energy and angular momentum from
the BH to the hot spot as follows:

\begin{equation}
\label{eq19} P_{MC}^{HS} = \left( {\lambda - 1}
\right)P_{MC}^{axis} + \varepsilon P_{MC}^{axis} = \left( {\lambda
+ \varepsilon - 1} \right)P_{MC}^{axis} ,
\end{equation}

\begin{equation}
\label{eq20} T_{MC}^{HS} = \left( {\lambda - 1}
\right)T_{MC}^{axis} + \varepsilon T_{MC}^{axis} = \left( {\lambda
+ \varepsilon - 1} \right)T_{MC}^{axis} .
\end{equation}

\noindent From equations (\ref{eq15}), (\ref{eq16}), (\ref{eq19})
and (\ref{eq20}) we find that the ratio of $P_{MC}^{HS} $ to
$P_{MC}^{naxis} $ is the same as that of $T_{MC}^{HS} $ to
$T_{MC}^{naxis} $, i.e.

\begin{equation}
\label{eq21} \begin{array}{l} R\left( {\delta ,\varepsilon }
\right) = \frac{P_{MC}^{HS} }{P_{MC}^{naxis} } = \frac{T_{MC}^{HS}
}{T_{MC}^{naxis} }  = \frac{\lambda + \varepsilon - 1}{\lambda }
\\ \\\quad\quad\quad\quad= 1 - \frac{1 - \varepsilon }{\left( {1 + \delta
\varepsilon } \right)^2},
\end{array}
\end{equation}

\noindent where the ratio $R\left( {\delta ,\varepsilon } \right)$
is used to express the fractional energy and angular momentum
attributable to the hot spot in the MC process. The curves of
$R\left( {\delta ,\varepsilon } \right)$ varying with the
parameters $\delta $ and $\varepsilon $ are shown in Fig.4.

From Fig.4 we find that the ratio $R\left( {\delta ,\varepsilon }
\right)$ is independent of the parameters $M$, $a_ * $ and $n$,
and it increases almost linearly with the parameters $\delta $ and
$\varepsilon $. For an appropriate angular region of the bulging
magnetic field on the horizon, such as $\varepsilon = 0.1$, we
have the ratio varying from 0.1 to 0.18 for a small value range of
the strength of the bulging magnetic field,\textbf{ }$0 < \delta <
0.5$\textbf{. }These results show that the effects of the bulging
magnetic field  on the energy and angular momentum transferred to
the hot spot are significant.

\begin{figure}
\begin{center}
{\includegraphics[width=6cm]{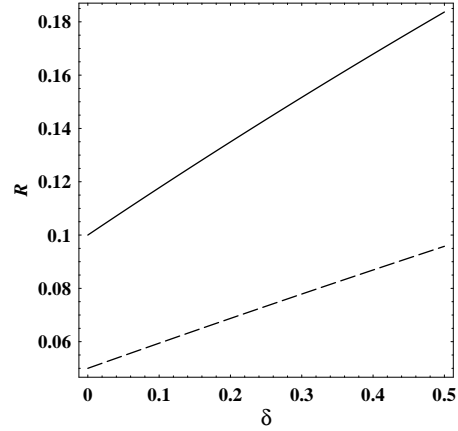}
 \centerline{\hspace{1cm}(a)}
 \includegraphics[width=6cm]{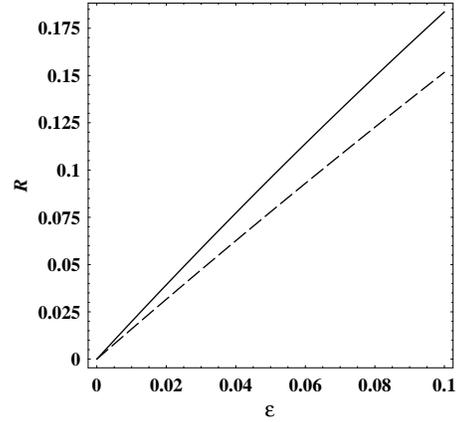}
 \centerline{\hspace{1cm}(b)}}
 \label{fig4}
 \caption{The curves of $R(\delta,\varepsilon)$ (a) versus $\delta$ for $0<\delta<0.5$ with $\varepsilon=0.1,0.05$ and (b)
 versus $\varepsilon$ for $0<\varepsilon<0.1$ with $\delta=0.5,0.3$ in solid and dashed lines, respectively.}
\end{center}
\end{figure}

Since the hot spot is located very near to the inner edge of the
disc, it should be optically thick. So we can define its effective
radiation temperature $\left( {T_{HS} } \right)_{eff} $ as a
blackbody spectrum:

\begin{equation}
\label{eq22} \left( {T_{HS} } \right)_{eff} = \left[ {{\left(
{F_{DA} + F_{MC}^{HS} } \right)} \mathord{\left/ {\vphantom
{{\left( {F_{DA} + F_{MC}^{HS} } \right)} \sigma }} \right.
\kern-\nulldelimiterspace} \sigma } \right]^{1 \mathord{\left/
{\vphantom {1 4}} \right. \kern-\nulldelimiterspace} 4}
\end{equation}

\noindent where $\sigma $ is the Stefan-Boltzmann constant. The
fluxes $F_{DA} $ and $F_{MC}^{HS} $ are radiated from the hot spot
due to disc accretion and the MC process, respectively, and are
expressed as follows (Page {\&} Thorne 1974; Li02a):

\begin{equation}
\label{eq23}
 \begin{array}{l}
  F_{DA} = - \frac{\dot {M}_D }{4\pi M^2\chi
_{ms}^4 }\frac{d\Omega _D }{\xi d\xi }\times \\ \\
\quad\quad\quad\quad\left( {E^ + - \Omega _D L^ + } \right)^{ -
2}\int_1^\xi {\left( {E^ + - \Omega _D L^ + } \right)\frac{dL^ +
}{d\xi }d\xi } ,
\end{array}
\end{equation}

\begin{equation}
\label{eq24} \begin{array}{l}
 F_{MC}^{HS} = - \frac{d\Omega _D
}{\xi d\xi }\left( {E^ + - \Omega _D L^ + } \right)^{ - 2}\times
\\ \\
\quad\quad\quad\quad\quad\quad\quad\int_1^\xi {\left( {E^ + -
\Omega _D L^ + } \right)H_{HS} \xi d\xi } .
 \end{array}
\end{equation}

\begin{figure*}
\label{fig5}
\begin{center}
{\includegraphics[width=5.6cm]{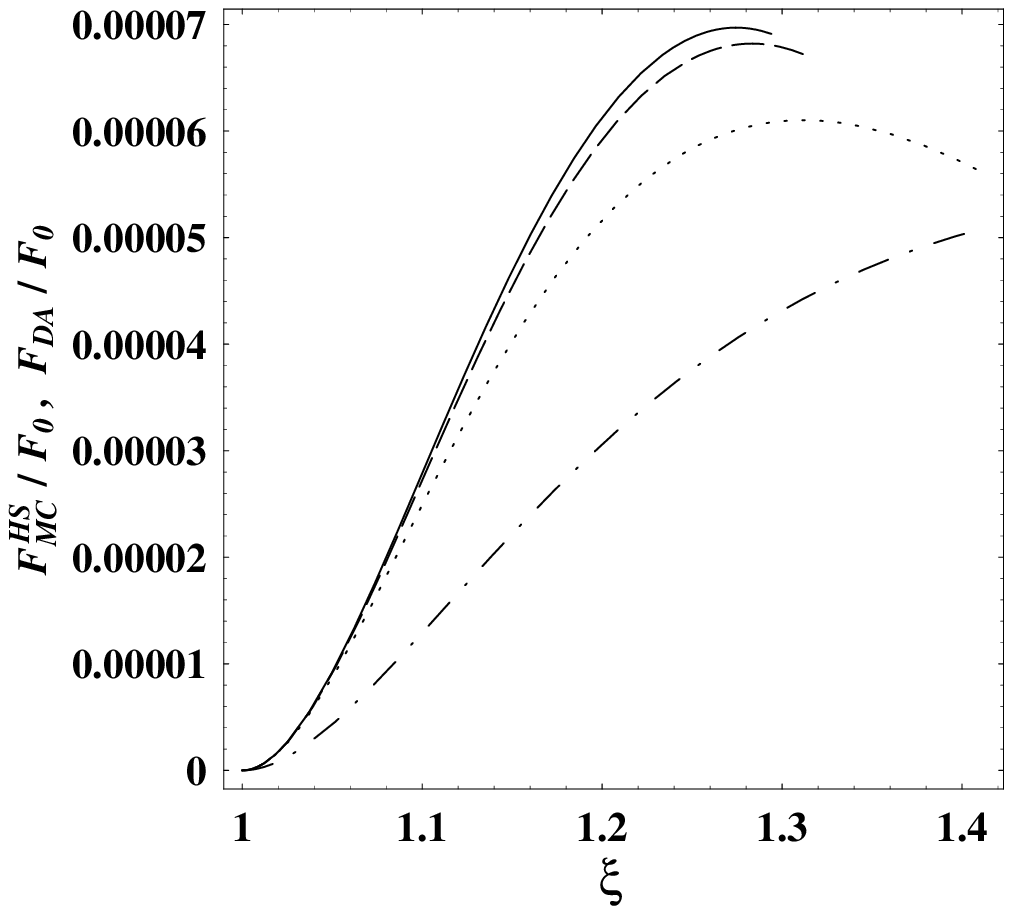}
 \includegraphics[width=5.6cm]{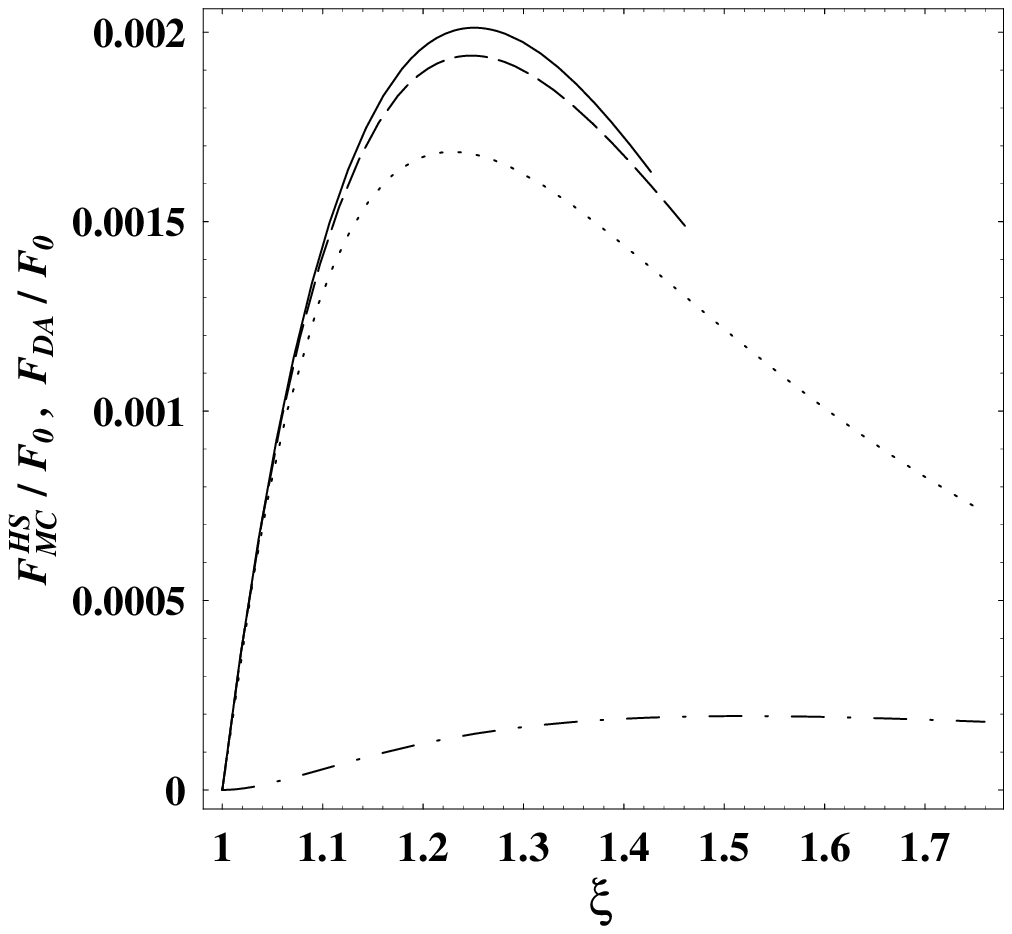}
  \includegraphics[width=5.6cm]{f5b.eps}
  \centerline{\hspace{0.6cm}(a)\hspace{5.5cm}(b)\hspace{5.6cm}(c)}}
\caption{The curves of ${F_{DA} } /{F_0 }$ (dot-dashed lines) and
${F_{MC}^{HS} } /{F_0 }$ for $1 < \xi < \xi _{out} $ and
$\varepsilon = 0.1$, $\delta = 0.5$ with $n = 1.1, 1.5$ and $3.0$
in solid, dashed and dotted lines, respectively. (a) $a_ * =
0.3594$; (b) $a_* = 0.80$; (c) $a_* = 0.998$.}
\end{center}
\end{figure*}

In this paper we only discuss the hot spot produced by the bulging
magnetic field on the BH with spin greater than $a_ * = 0.3594$,
which corresponds to \textit{case II} in equation (\ref{eq10}). In
equations (\ref{eq23}) and (\ref{eq24}) $E^ + $ and $L^ + $ are
specific energy and angular momentum of accreting particles,
respectively, and are expressed respectively by (Novikov {\&}
Thorne 1973)

\begin{equation}
\label{eq25} E^ +
=\frac{1-2\chi^{-2}+a_*\chi^{-3}}{(1-3\chi^{-2}+2a_*\chi^{-3})^{1/2}}
\end{equation}

\noindent and

\begin{equation}
\label{eq26} L^ + =  \frac{M\chi(1 - 2a_ * \chi ^{ - 3} + a_ * ^2
\chi ^{ - 4})}{(1-3\chi^{-2}+2a_*\chi^{-3})^{1/2}}.
\end{equation}

In equation (\ref{eq24}) $H_{HS} $ is the flux of angular momentum
transferred from the BH to the hot spot by the bulging magnetic
field, being related to $T_{MC}^{HS} $ by

\begin{equation}
\label{eq27} 4\pi \varepsilon rH_{HS} = 2r\Delta \phi H_{HS} =
\partial {T_{MC}^{HS} } \mathord{\left/ {\vphantom {{T_{MC}^{HS} }
{\partial r}}} \right. \kern-\nulldelimiterspace} {\partial r}
\end{equation}

\noindent Combining equations (\ref{eq2}), (\ref{eq3}),
(\ref{eq14}), (\ref{eq20}) and (\ref{eq27}), we have

\begin{equation}
\label{eq28} {H_{HS} } \mathord{\left/ {\vphantom {{H_{HS} } {H_0
}}} \right. \kern-\nulldelimiterspace} {H_0 } = \left\{
{\begin{array}{l}
 A\left( {a_ * ,\xi ,n,\delta ,\varepsilon } \right)\xi ^{ - n}, 1 <
\xi<\xi _{out} \\
 \quad\quad 0,\quad\quad\quad\quad\quad\quad\mbox{ }\xi > \xi _{out}  \\
 \end{array},} \right.
\end{equation}

\noindent where

\begin{equation}
\label{eq29} \left\{ {\begin{array}{l}
 A\left( {a_ * ,\xi ,n,\delta ,\varepsilon } \right) = \frac{\left( {1 +
2\delta + \delta ^2\varepsilon } \right)a_ * \left( {1 - \beta }
\right)\left( {1 + q} \right)}{2\pi \chi _{ms}^2 \left[ {2\csc
^2\theta - \left( {1 - q} \right)} \right]}F_A \left( {a_ * ,\xi }
\right) \\ \\
 F_A \left( {a_ * ,\xi } \right) =\\
 \quad\quad\quad \frac{\sqrt {1 + a_ * ^2 \chi _{ms}^{ -
4} \xi ^{ - 2} + 2a_ * ^2 \chi _{ms}^{ - 6} \xi ^{ - 3}} }{\sqrt
{\left( {1 + a_ * ^2 \chi _{ms}^{ - 4} + 2a_ * ^2 \chi _{ms}^{ -
6} } \right)\left( {1 - 2\chi _{ms}^{ - 2} \xi ^{ - 1} + a_ * ^2
\chi _{ms}^{ - 4} \xi ^{ - 2}}
\right)} } \\
 \end{array}} \right.
\end{equation}

\noindent and

\begin{equation}
\label{eq30} H_0 = \left\langle {B_H^2 } \right\rangle M =
1.48\times 10^{13}\times B_4^2 m_{BH} \mbox{ }g \cdot s^{ - 2}.
\end{equation}

\noindent In equation (\ref{eq30}) $B_4 $ and $m_{BH} $ are $\sqrt
{\left\langle {B_H^2 } \right\rangle } $ and $M$ in the units of
$10^4Gauss$ and one solar mass, respectively.

Since the magnetic field on the horizon is brought and held by its
surrounding magnetized disc (Macdonald {\&} Thorne 1982; Thorne,
Press {\&} Macdonald 1986), there must exist some relations
between $B_H $ and $\dot {M}_D $. As a matter of fact these
relations might be rather complicated, and would be very different
in different situations. One of them is given by considering the
balance between the pressure of the magnetic field on the horizon
and the ram pressure of the innermost parts of an accretion flow
(Moderski, Sikora {\&} Lasota 1997), i.e.

\begin{equation}
\label{eq31} {B_H^2 } \mathord{\left/ {\vphantom {{B_H^2 } {\left(
{8\pi } \right)}}} \right. \kern-\nulldelimiterspace} {\left(
{8\pi } \right)} = P_{ram} \sim \rho c^2\sim {\dot {M}_D }
\mathord{\left/ {\vphantom {{\dot {M}_D } {\left( {4\pi r_H^2 }
\right)}}} \right. \kern-\nulldelimiterspace} {\left( {4\pi r_H^2
} \right)},
\end{equation}

\noindent From equation (\ref{eq31}) we assume the relation as

\begin{equation}
\label{eq32} \dot {M}_D = {\left\langle {B_H^2 } \right\rangle
M^2\left( {1 + q} \right)^2} \mathord{\left/ {\vphantom
{{\left\langle {B_H^2 } \right\rangle M^2\left( {1 + q} \right)^2}
2}} \right. \kern-\nulldelimiterspace} 2.
\end{equation}

The accretion rate $\dot {M}_D $ is still axisymmetric in spite of
the existence of the bulging magnetic field on the horizon,
provided that the extra angular momentum transferred into the disc
is totally radiated away from the disc. Incorporating equations
(\ref{eq32}), (\ref{eq23}), (\ref{eq24}) and (\ref{eq28}), we have

\begin{equation}
\label{eq33} \left\{ {\begin{array}{l}
 {F_{DA} } \mathord{\left/ {\vphantom {{F_{DA} } {F_0 }}} \right.
\kern-\nulldelimiterspace} {F_0 } = f_{DA} \left( {a_ * ,\xi } \right)
\\ \\
 f_{DA} \left( {a_ * ,\xi } \right) = - \frac{\left( {1 + q} \right)^2}{8\pi
\chi _{ms}^4 \left( {E^ + - \Omega _D L^ + } \right)^2}\times\\
\\\quad\quad\quad\quad\quad\quad\quad\quad\quad\frac{d\Omega _D }{\xi
d\xi }\int_1^\xi {\left( {E^ + - \Omega _D L^ + } \right)\frac{dL^
+
}{d\xi }d\xi } \\
 \end{array}} \right.
\end{equation}

\noindent and

\begin{equation}
\label{eq34} \left\{ {\begin{array}{l}
 {F_{MC}^{HS} } \mathord{\left/ {\vphantom {{F_{MC}^{HS} } {F_0 }}} \right.
\kern-\nulldelimiterspace} {F_0 } = f_{MC} , \\ \\
 f_{MC} = - \frac{1}{\left( {E^ + - \Omega _D L^ + }
\right)^2}\frac{Md\Omega _D }{\xi d\xi }\times\\
\\\quad\quad\quad\quad\int_1^\xi {\left( {E^ + - \Omega _D L^ + } \right)A\left(
{a_ * ,\xi ,n,\delta ,\varepsilon } \right)\xi ^{1 -
n}d\xi ,} \\
 \end{array}} \right.
\end{equation}

\noindent with

\begin{equation}
\label{eq35} F_0 = \left\langle {B_H^2 } \right\rangle c =
2.998\times 10^{18}B_4^2 \mbox{ }erg \cdot cm^{ - 2} \cdot s^{ -
1}.
\end{equation}

\begin{table*}
\caption{Radiation energy and frequency of a rotating hot spot
produced by the MC process and disc accretion with $B_4 = 1$,
$\delta = 0.5$ and $\varepsilon = 0.2$.}
\begin{tabular}
{|p{39pt}|p{39pt}|p{86pt}|p{86pt}|p{86pt}|p{86pt}|} \hline $a_ *
$& $n$& $\left( {T_{HS}^{\max } } \right)_{eff} $(K)&
$E_{HS}^{\max } $(kev)& $\xi _{\max } $&
$m_{BH} \nu _{HS} $(Hz) \\
\hline \raisebox{-3.00ex}[0cm][0cm]{0.6}& 1.1& 7.722$\times 10^4$&
6.661$\times 10^{ - 3}$& 1.348&
2620.37 \\

 &
1.5& 7.666$\times 10^4$& 6.612$\times 10^{ - 3}$& 1.367&
2569.23 \\
&
 3.0& 7.440$\times 10^4$& 6.417$\times 10^{ - 3}$& 1.447&
2366.91 \\
\hline \raisebox{-3.00ex}[0cm][0cm]{0.8}& 1.1& 1.038$\times 10^5$&
8.953$\times 10^{ - 3}$& 1.380&
3657.66 \\

 &
1.5& 1.029$\times 10^5$& 8.875$\times 10^{ - 3}$& 1.396&
3599.78 \\

 &
3.0& 9.950$\times 10^4$& 8.582$\times 10^{ - 3}$& 1.438&
3455.97 \\
\hline \raisebox{-3.00ex}[0cm][0cm]{0.998}& 1.1& 1.994$\times
10^5$& 1.720$\times 10^{ - 2}$& 1.250&
11061.7 \\

 &
1.5& 1.982$\times 10^5$& 1.709$\times 10^{ - 2}$& 1.247&
11084.3 \\

 &
3.0& 1.937$\times 10^5$& 1.671$\times 10^{ - 2}$& 1.227&
11261.7 \\
\hline
\end{tabular}
 \label{tab1}
\end{table*}

Inspecting equations (\ref{eq33})---(\ref{eq35}), we obtain the
following results:

(i) Both $F_{DA} $ and $F_{MC}^{HS} $ are proportional to
$\left\langle {B_H^2 } \right\rangle $, and are independent of the
BH mass due to equation (\ref{eq32}).

(ii) The bulging magnetic field on the horizon only affects
$F_{MC}^{HS} $ rather than $F_{DA} $, since $\dot {M}_D $ is
independent of the bulging magnetic field.

By using equations (\ref{eq33}) and (\ref{eq34}) we have the
curves of $F_{DA} $ and $F_{MC}^{HS} $ varying with $\xi $ as
shown in Fig. 5.

From Fig.5 we obtain the following results:

(i) The relation $F_{MC}^{HS} > F_{DA} $ always holds for $0.3594
< a_ * < 1$, implying that the contribution from the MC process
always dominates over that from disc accretion for the above value
range of the BH spin.

(ii) Both $F_{MC}^{HS} $ and $F_{DA} $ approach zero as $\xi \to
\xi _{in} = 1$, and both vary non-monotonically with $\xi $. The
flux $F_{MC}^{HS} $ varies much more sharply than $F_{DA} $ does,
attaining its peak value between the inner and outer boundaries of
the hot spot.

(iii) The peak of $F_{MC}^{HS} $ is closer to the inner edge of
the disc than that of $F_{DA} $. The more is the value of $a_ * $,
the greater is the peak value, and the closer is the peak to the
inner edge.

Substituting equations (\ref{eq33}) and (\ref{eq34}) into equation
(\ref{eq22}), we have the effective temperature and the
corresponding radiation energy of the hot spot as follows:

\begin{equation}
\label{eq36} \left( {T_{HS} } \right)_{eff} = T_0 \left[ {f_{DA} +
f_{MC} } \right]^{1 \mathord{\left/ {\vphantom {1 4}} \right.
\kern-\nulldelimiterspace} 4},
\end{equation}

\begin{equation}
\label{eq37} E_{HS} \equiv k_B \left( {T_{HS} } \right)_{eff} =
E_0 \left[ {f_{DA} + f_{MC} } \right]^{1 \mathord{\left/
{\vphantom {1 4}} \right. \kern-\nulldelimiterspace} 4},
\end{equation}

\noindent where

\begin{equation}
\label{eq38} T_0 = \left( {{F_0 } \mathord{\left/ {\vphantom {{F_0
} \sigma }} \right. \kern-\nulldelimiterspace} \sigma } \right)^{1
\mathord{\left/ {\vphantom {1 4}} \right.
\kern-\nulldelimiterspace} 4} \approx 4.8\times 10^5B_4^{1
\mathord{\left/ {\vphantom {1 2}} \right.
\kern-\nulldelimiterspace} 2} K,
\end{equation}

\noindent and

\begin{equation}
\label{eq39} E_0 = k_B \left( {{F_0 } \mathord{\left/ {\vphantom
{{F_0 } \sigma }} \right. \kern-\nulldelimiterspace} \sigma }
\right)^{1 \mathord{\left/ {\vphantom {1 4}} \right.
\kern-\nulldelimiterspace} 4} \approx 4.14\times 10^{ - 2}B_4^{1
\mathord{\left/ {\vphantom {1 2}} \right.
\kern-\nulldelimiterspace} 2} kev.
\end{equation}

\noindent The maxima of $\left( {T_{HS} } \right)_{eff} $ and
$E_{HS} $ corresponding to the different values of the BH spin and
the power-law index are shown in Table 1.

It is easy to prove that the maxima of either $\left( {T_{HS} }
\right)_{eff} $ or $E_{HS} $ increase monotonically with the
increasing BH spin. For the fast spinning BH with spin $0.6 < a_ *
< 0.998$ we conclude that the radiation energy of the hot spot
reaches $0.006 \sim 0.017kev$, provided that the bulging magnetic
field reaches $B_4 $ or $10^4Gauss$, while this value range
becomes $0.6 \sim 1.7kev$ for the bulging magnetic field reaching
$10^4B_4 $ or $10^8Gauss$. So the hot spot can have energy
reaching soft to hard X-ray in the MC process, provided that a
fast-spinning BH has the bulging magnetic field of $10^4 \sim
10^8Gauss$ on the horizon.

Since the magnetic field lines are frozen in the disc plasma, the
rotating frequency of the hot spot is regarded as QPO frequency.
As suggested by Nowak and Lehr (1998), the QPO frequency can be
worked out by calculating Keplerian angular velocity at the place
where $r^2F$ attains its maximum, and $F$ is the sum of $F_{MC} $
and $F_{DA} $. We can define a dimensionless function as follows:

\begin{equation}
\label{eq40} F_{QPO} \equiv {r^2F} \mathord{\left/ {\vphantom
{{r^2F} {r_{ms}^2 }}} \right. \kern-\nulldelimiterspace} {r_{ms}^2
}F_0 = \xi ^2\left( {{F_{DA} } \mathord{\left/ {\vphantom {{F_{DA}
} {F_0 }}} \right. \kern-\nulldelimiterspace} {F_0 } +
{F_{MC}^{HS} } \mathord{\left/ {\vphantom {{F_{MC}^{HS} } {F_0 }}}
\right. \kern-\nulldelimiterspace} {F_0 }} \right)
\end{equation}

\begin{figure}
\begin{center}
\includegraphics[width=6cm]{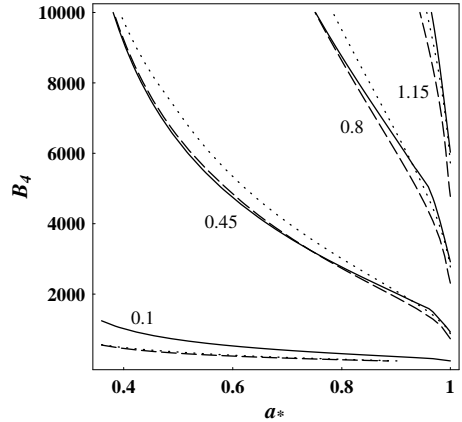}
\caption{The contours of $E_{HS} $ of constant values (kev) for
$0.3594 < a_ * < 0.998$ and $10^3 < B_4 < 10^4$ with $\delta =
0.5$, $\varepsilon = 0.1$ and $n = 1.1, 1.5$, $3.0$ in solid,
dashed and dotted lines, respectively.}
 \label{fig6}
 \end{center}
\end{figure}

\begin{figure}
\begin{center}
\includegraphics[width=6cm]{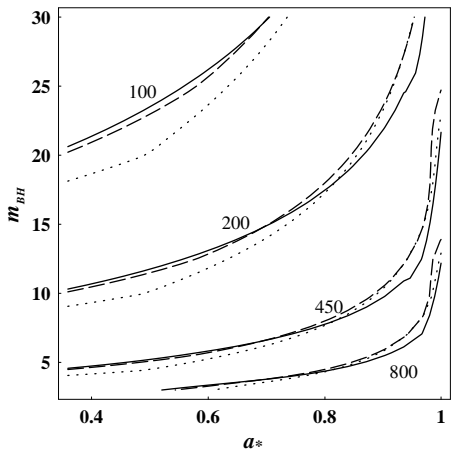}
\caption{The contours of $\nu _{HS} $ of constant values (Hz) for
$0.3594 < a_ * < 0.998$ and $3 < m_{BH} < 30$ with $\delta = 0.5$,
$\varepsilon = 0.1$ and $n = 1.1, 1.5$, $3.0$ in solid, dashed and
dotted lines, respectively.}
 \label{fig7}
 \end{center}
\end{figure}

\noindent Incorporating equations (\ref{eq33}), (\ref{eq34}) and
(\ref{eq40}), we obtain $\xi _{\max } $ corresponding to the
maximum of function $F_{QPO} $, and the QPO frequency $\nu _{HS} $
can be calculated by

\begin{equation}
\label{eq41} \nu _{HS} \equiv \nu _0 (\xi _{\max }^{3
\mathord{\left/ {\vphantom {3 2}} \right.
\kern-\nulldelimiterspace} 2} \chi _{ms}^3 + a_ * )^{ - 1},
\end{equation}

\noindent where we have $\nu _{HS} \equiv \left. {{\Omega _D }
\mathord{\left/ {\vphantom {{\Omega _D } {2\pi }}} \right.
\kern-\nulldelimiterspace} {2\pi }} \right|_{\xi = \xi _{\max } }
$ and $\nu _0 \equiv \left( {m_{BH} } \right)^{ - 1}\times
3.23\times 10^4Hz$. The values of $\xi _{\max } $ and $\nu _{HS} $
corresponding to different values of BH spin and power-law index
are shown in Table 1, where $\xi _{\max } $ turns out to be very
close to the inner edge of the disc. For the fast-spinning BH with
spin $0.6 < a_
* < 0.998$ we have $\nu _{HS} $ varying from ${ \sim 2.4kHz} \mathord{\left/
{\vphantom {{ \sim 2.4kHz} {m_{BH} }}} \right.
\kern-\nulldelimiterspace} {m_{BH} }$ to ${ \sim 11kHz}
\mathord{\left/ {\vphantom {{ \sim 11kHz} {m_{BH} }}} \right.
\kern-\nulldelimiterspace} {m_{BH} }$ with $m_{BH} > 3$.

In order to describe the variation of $E_{HS} $ and $\nu _{HS} $
of the rotating hot spot more clearly we plot the contours of
$E_{HS} $ and $\nu _{HS} $ in terms of the concerning parameters
as shown in Figs 6 and 7, respectively.

From Fig. 6 we find that less magnetic field is required as the BH
spin increases for the given value of $E_{HS} $. Furthermore the
more is the BH spin, the more steeply the contours of $E_{HS} $
decline. This result implies that the MC process becomes very
efficient in transferring energy from a very fast spinning BH to
the hot spot.

From Fig. 7 we find that more BH mass is required as the BH spin
increases for the given values of $\nu _{HS} $. Furthermore the
more is the BH spin, the more contours of $\nu _{HS} $ will be
crossed over for the given BH mass. This result implies that $\nu
_{HS} $ increases very rapidly as the BH spin $a_ * $ approaches
unity for the given BH mass.

The BH spin acts as a bridge to link the contours of $E_{HS} $ and
$\nu _{HS} $ in Figs 6 and 7. This linkage is helpful to fit
observations of kHz QPOs in X-ray binaries in some scope.

\section{ FLUCTUATION OF PARAMETERS AND WIDTHS OF QPO FREQUENCIES}

It is shown that the widths of observed frequencies of kilohertz
QPOs in X-ray binaries are rather narrow, usually varying from
several to some dozens of hertz, e.g. as shown in Tables 1-3 given
by van der Klis (2000). However, it is difficult for us to explain
the narrow range of QPOs by the rotating hot spot produced by the
stationary bulging magnetic field on the horizon.

As a matter of fact we can fit the narrow range of QPO widths by
the fluctuation of $\delta $ and $\varepsilon $ in our model. The
argument is given as follows. From equation (\ref{eq41}) we find
that QPO frequency is determined by $\xi _{\max } $, and the
latter depends on the maximum of function $F_{QPO} $ in equation
(\ref{eq40}), which consists of the sum of ${F_{DA} }
\mathord{\left/ {\vphantom {{F_{DA} } {F_0 }}} \right.
\kern-\nulldelimiterspace} {F_0 }$ and ${F_{MC}^{HS} }
\mathord{\left/ {\vphantom {{F_{MC}^{HS} } {F_0 }}} \right.
\kern-\nulldelimiterspace} {F_0 }$. As mentioned above that the
peak of ${F_{MC}^{HS} } \mathord{\left/ {\vphantom {{F_{MC}^{HS} }
{F_0 }}} \right. \kern-\nulldelimiterspace} {F_0 }$ is located
closer to the inner edge of the disc than that of ${F_{DA} }
\mathord{\left/ {\vphantom {{F_{DA} } {F_0 }}} \right.
\kern-\nulldelimiterspace} {F_0 }$, so we infer that the narrow
range of QPO widths might be produced by the variation of
${F_{MC}^{HS} } \mathord{\left/ {\vphantom {{F_{MC}^{HS} } {F_0
}}} \right. \kern-\nulldelimiterspace} {F_0 }$, provided that the
parameters $\delta $ and $\varepsilon $ are fluctuating rather
than constant. The curves of ${d\nu _{HS} } \mathord{\left/
{\vphantom {{d\nu _{HS} } {\nu _{HS} }}} \right.
\kern-\nulldelimiterspace} {\nu _{HS} }$ versus $a_ * $ for the
given fluctuation of $\delta $ and $\varepsilon $ are shown in
Fig.8.

\begin{figure}
\begin{center}
{\includegraphics[width=6cm]{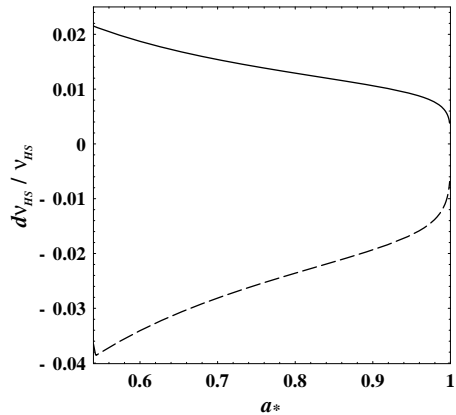}
 \centerline{\hspace{1cm}(a)}
 \includegraphics[width=6cm]{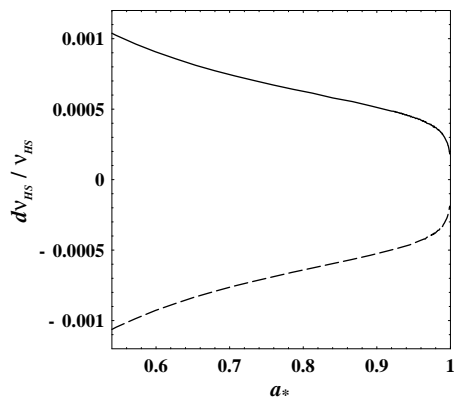}
 \centerline{\hspace{1.2cm}(b)}}
\caption{The curves of ${d\nu _{HS} } / {\nu _{HS} } $ versus $a_
* $ for $0.54 < a_ * < 0.998$, (a) $n = 3$, $\varepsilon = 0.2$,
$\Delta \varepsilon = 0$, $\delta = 05$, $\Delta \delta = 0.3$
(solid line), $\Delta \delta = - 0.3$ (dashed line), (b) $n = 3$,
$\delta = 0.5$, $\Delta \delta = 0$, $\varepsilon = 0.2$, $\Delta
\varepsilon = 0.1$ (solid line), $\Delta \varepsilon = - 0.1$
(dashed line).} \label{fig8}
\end{center}
\end{figure}

From Fig.8 we have the following results.

(i) The value of ${d\nu _{HS} } \mathord{\left/ {\vphantom {{d\nu
_{HS} } {\nu _{HS} }}} \right. \kern-\nulldelimiterspace} {\nu
_{HS} }$ varies between --0.04 to 0.02 with $\left| {\Delta \delta
} \right| = 0.3$ and $\Delta \varepsilon = 0$ for $0.54 < a_ * <
0.998$. This result implies that the range of QPO widths is of
dozens of hertz, and could be produced by the fluctuation of the
strength of the bulging magnetic field of the BH as shown in
Fig.8(a).

(ii) The effect of the variation of $\varepsilon $ on the range of
kHz QPO width is about one order of magnitude less than that of
$\delta $ as shown in Fig.8(b).

The values of ${d\nu _{HS} } \mathord{\left/ {\vphantom {{d\nu
_{HS} } {\nu _{HS} }}} \right. \kern-\nulldelimiterspace} {\nu
_{HS} }$ corresponding to different values of the concerning
parameters are listed in Table 2. As shown in Fig.8(a) and Table 2
we can fit the range of kHz QPO width by a variety of parameters
in our model, such as $a_\ast $, $n$, $\varepsilon $, $\delta $,
$\Delta \varepsilon $ and $\Delta \delta $.


\begin{table*}
 \caption{ The values of ${d\nu _{HS} } /{\nu _{HS} } $ corresponding to
different values of the concerning parameters with $\varepsilon =
0.2$, $\Delta \varepsilon = 0$ and  $\delta = 0.5$.}
\begin{tabular}
{|p{36pt}|p{31pt}|p{47pt}|p{77pt}|p{77pt}|p{77pt}|p{77pt}|} \hline
\raisebox{-1.50ex}[0cm][0cm]{$a_\ast $}&
\raisebox{-1.50ex}[0cm][0cm]{$n$}&
\raisebox{-1.50ex}[0cm][0cm]{$m_{BH} \nu _{HS} $}&
\multicolumn{4}{|p{310pt}|}{\hspace{4.8cm}${d\nu _{HS} } / {\nu _{HS} }$}  \\
\cline{4-7}\\
 &
 &
 &
$\Delta \delta = 0.1$ & $\Delta \delta = - 0.1$ & $\Delta \delta =
0.2$ &
$\Delta \delta = - 0.2$  \\
\hline \raisebox{-3.00ex}[0cm][0cm]{0.600}& 1.1& 2620.37&
2.42$\times 10^{ - 3}$ &\hspace{0.5cm}
--- &
4.46$\times 10^{ - 3}$ &\hspace{0.5cm}
---  \\

 &
1.5& 2569.23& 3.01$\times 10^{ - 3}$ &-3.63$\times 10^{ - 3}$ &
5.55$\times 10^{ - 3}$ &
-6.36$\times 10^{ - 3}$  \\

 &
3.0& 2366.91& 7.35$\times 10^{ - 3}$ & -8.93$\times 10^{ - 3}$ &
13.5$\times 10^{ - 3}$ &
-20.0$\times 10^{ - 3}$  \\
\hline \raisebox{-3.00ex}[0cm][0cm]{0.800}& 1.1& 3657.66&
1.94$\times 10^{ - 3}$ & -2.34$\times 10^{ - 3}$ & 3.57$\times
10^{ - 3}$ &
-5.23$\times 10^{ - 3}$  \\

 &
1.5& 3599.78& 2.41$\times 10^{ - 3}$ & -2.91$\times 10^{ - 3}$ &
4.43$\times 10^{ - 3}$ &
-6.50$\times 10^{ - 3}$  \\

 &
3.0& 3455.97& 5.07$\times 10^{ - 3}$ & -6.17$\times 10^{ - 3}$ &
9.32$\times 10^{ - 3}$ &
-13.8$\times 10^{ - 3}$  \\
\hline \raisebox{-3.00ex}[0cm][0cm]{0.998}& 1.1& 11061.7&
1.31$\times 10^{ - 3}$ & -1.58$\times 10^{ - 3}$ & 2.42$\times
10^{ - 3}$ &
-3.53$\times 10^{ - 3}$  \\

 &
1.5& 11084.3& 1.44$\times 10^{ - 3}$ & -1.74$\times 10^{ - 3}$ &
2.65$\times 10^{ - 3}$ &
-3.87$\times 10^{ - 3}$  \\

 &
3.0& 11261.7& 1.78$\times 10^{ - 3}$ & -2.15$\times 10^{ - 3}$ &
3.27$\times 10^{ - 3}$ &
-4.81$\times 10^{ - 3}$  \\
\hline
\end{tabular}

 \label{tab2}
\end{table*}

\section{ SUMMARY}

In this paper an analytic model of rotating hot spot is proposed
to explain kHz QPOs in X-ray binaries by using the MC of a
rotating BH with the disc. Six parameters are used in our model.

(i) The parameters $\delta $, $\varepsilon $ and $B_4 $ are given
for describing the bulging magnetic field on the horizon by
equation (\ref{eq1}). Specifically speaking, the parameter $B_4 $
is used to indicate root-mean-square of the magnetic field over
the poloidal angular coordinate from $\theta _1 $ to $\theta _2 $.
The parameters $\delta $ and $\varepsilon $ are in charge of the
fractional energy and angular momentum attributable to the hot
spot by equation (\ref{eq21}). The fluctuation of $\delta $ and
$\varepsilon $ can affect the range of kHz QPO widths.

(ii) The parameters $m_{BH} $ and $a_ * $ are given for describing
the mass and spin of the BH, which determine the frequency of QPOs
by equations (\ref{eq41}).

(iii) The parameter $n$ is used to indicate the concentration of
magnetic field on the central region of the disc, and it is
involved in equation (\ref{eq7}) to confine the radial width of
the hot spot. In addition, it is involved in equation (\ref{eq40})
to determine $\xi _{\max } $. We can find that the index $n$ plays
a role of fine-tuning energy and frequency of the hot spot and the
range of kHz QPOs widths as shown in Table 1, Fig.6, Fig.7 and
Table 2, respectively. \\

\noindent\textbf{Acknowledgments. }This work is supported by the
National Natural Science Foundation of China under Grant No.
10173004 and No. 10121503. The anonymous referee is thanked for
his suggestion about the effects of the parameters $\delta $ and
$\varepsilon $ on the rotating hot spot.

\end{document}